\documentstyle[titlepage,12pt]{article}
\begin{document}
\parindent 2em

\begin{titlepage}
\begin{center}
\vspace{12mm}
{\LARGE Continuum dual theory of the transition in 3D 
lattice superconductor}  
\vspace{15mm}

Igor F. Herbut\\

Department of Physics and Astronomy, University of British Columbia, 
6224 Agricultural Road, Vancouver B. C., Canada V6T 1Z1\\

\end{center}
\vspace{10mm}
{\bf Abstract:}
A recently proposed form of dual theory for three dimensional 
superconductor is rederived starting from the lattice electrodynamics 
and studied by renormalization group.
The superfluid density below and close to the transition vanishes as 
inverse of the correlation length for the disorder field. The 
corresponding universal amplitude is given by the fixed point value 
of the dual charge, and it is calculated to the leading order. 
The continuum dual theory predicts the divergence of the 
magnetic field penetration depth with the XY exponent, in contradiction 
to the results obtained from the Ginzburg-Landau 
theory for the superconducting 
order-parameter. Possible reasons for this difference are discussed.


\end{titlepage}

%\section{ }

Duality transformations in statistical mechanics have 
led to 
important insights into the nature of phase transitions in various 
systems, ever since the discovery of self-duality in 
two-dimensional (2D) Ising model by Kramers and Wannier \cite{kramers}.
This concept has also proved important 
in studies of the phase transition in 2D 
XY model, where dual partition function factorizes into 
spin-wave and vortex contributions \cite{villain}, 
and therefore provides a natural 
framework for description of 
the Kosterlitz-Thouless transition.  
Similarly, the phase transition in 3D XY model 
can be understood within a dual approach, where the basic objects,  
instead of vortices, now become vortex loops. Critical 
temperature is identified as the point of 
proliferation of vortex loops through the system. 
The high-temperature phase of the original XY model corresponds to 
the low-temperature phase of the dual model, in which the disorder 
field that describes vortex loops has condensed \cite{book}. Critical 
exponent for the correlation length 
calculated within the dual approach \cite{cecio} 
is in very good agreement with the standard value, 
confirming the correctness of the physical picture of condensation of 
vortex loops at the transition in 3D XY model.

Critical behavior of superconductors is described by the Ginzburg-Landau 
 theory for a complex scalar field which, Cooper pairs being charged, is 
coupled to a fluctuating gauge potential. Nature of the phase transition 
in this system has long been a matter of debate: while close to four 
dimensions the transition is always first-order 
\cite{halperin}, \cite{coleman}, 
in three dimensions, 
it is believed that it 
can be both first and second order, depending on the value 
of the Ginzburg-Landau parameter $\kappa$ \cite{kleinert}, 
\cite{bartho}, \cite{herbut}, \cite{bergerhof}. 
While some  consensus has emerged 
on the overall topology of the phase diagram in the 3D case, the 
precise nature of the 
continuous transition in the theory is still controversial. 
Recent renormalization group 
analysis of the Ginzburg-Landau theory for superconducting 
order parameter yields a novel critical behavior, distinct from the 
one of the neutral superfluid \cite{herbut}, \cite{bergerhof}. 
Monte Carlo and analytical studies of 
the lattice version of the theory suggest, on the 
other hand, that the transition still belongs to the XY universality class 
\cite{dasgupta}, \cite{kleinert}. As a step towards a better understanding 
of these issues, in this paper the theory for 3D lattice superconductor 
is revisited. Under a certain plausible, but unproven assumption, 
the dual partition function of  
lattice superconductor may be expressed in a 
form that has a simple continuum limit. 
Based on different arguments, the  
same form of the continuum dual theory was previously proposed 
in ref. 12, and studied in ref. 13. 
However, there are some important 
differences between the assumptions and the conclusions of the 
present work and those of the previous study \cite{kiometzis}. 
The continuum dual 
model is analyzed by renormalization group in 3D and the transition 
described by this theory 
is confirmed to be in the universality class of 3D XY 
model. The stiffness constant proportional to the superfluid density 
vanishes as a power-law close to the transition, 
with a universal amplitude given by the 
fixed point value of the dual charge. In combination with 
the result on the anomalous dimension of the gauge field 
at the stable fixed point in the original theory \cite{herbut}, 
the continuum dual theory suggests that the magnetic field penetration depth 
diverges with the power $\nu_{xy}\simeq 2/3$. This agrees with the 
result of Peskin \cite{peskin}, but not with the conclusions 
of ref. 13. It also disagrees with the exponent derived 
from the Ginzburg-Landau 
theory \cite{herbut}, \cite{bergerhof}. Possible reasons for this 
difference are considered.

The partition function for lattice superconductor 
(3D lattice electrodynamics) is defined by \cite{peskin}, 
\cite{thomas}:
\begin{equation}
Z=\int'_A \int_\theta \exp(K_0 \sum\cos(\theta_{n+\hat{\mu}}-
\theta_{n}-A_{n\mu})-\frac{1}{2e^2}\sum (\nabla\times A)_{n\mu}^2 ), 
\end{equation}
where the sums run over sites and links of a 3D quadratic lattice, 
$\hat{\mu}=\hat{x},\hat{y}, \hat{z}$, 
$n\mu$ denotes an oriented link between 
$n$ and $n+\hat{\mu}$ sites, 
integrals over gauge potential in the transverse 
gauge $\nabla\cdot A=0$ run over whole real axis, integrals over phases 
are from $-\pi$ to $\pi$, and $\nabla\times A$ and $\nabla\cdot A$ 
are the lattice curl and the lattice divergence, respectively. 
$e$ is the charge of a Cooper pair and 
$K_0$ is the stiffness constant proportional to the 
superfluid density in the Meissner phase. 
The lattice superconductor corresponds to the extreme type-II 
limit of the Ginzburg-Landau model where the phase transition 
between normal and superconducting phases is expected to be continuous 
\cite{herbut}, \cite{bergerhof}.  
In the Villain 
approximation \cite{villain} the phases can be integrated out exactly, and 
the theory may be expressed in terms of 
integer-valued link variables $\{m_{n\mu}\}$ coupled to the 
gauge field:
\begin{equation}
Z=\int'_A \sum_{m}' \exp(-\frac{1}{2K_0} \sum m_{n\mu}^{2}+ 
i\sum m_{n\mu}
A_{n\mu} - \frac{1}{2e^2} \sum (\nabla\times A)_{n\mu}^2), 
\end{equation} 
where primes on the integral and on the sum, as throughout the paper, 
 denote the conditions 
 $\nabla\cdot A =\nabla\cdot m=0$. The second condition forces the integer 
variables $\{m_{n\mu}\}$ to form closed loops. For $e=0$ 
the gauge field decouples, and one is left with the standard 
loop  representation of 
the XY model, with the continuous phase transition at some $K_0=K_c$. 
If $K_0=\infty$, the sum over integers $\{m_{n\mu}\}$ can be 
performed, and the gauge-field becomes constrained to  
integer values. Remarkably, in this limit the lattice superconductor 
again maps onto XY model \cite{peskin}. 
Building on this observation, it has been argued that 
even for finite $K_0$ 
the lattice superconductor 
should exhibit the "inverted" XY transition \cite{dasgupta}. 
Here we  note that the partition function can be rewritten as 
\begin{equation}
Z=\int'_{A} \int'_{h} \sum_m \exp(-\frac{1}{2K_0}\sum 
(\nabla\times h)_{n\mu}^{2} 
+i\sum h_{n\mu} (\nabla \times A)_{n\mu} - 
\frac{1}{2 e^2} \sum (\nabla\times A)_{n\mu}^2 
+ i 2\pi \sum m_{n\mu} h_{n\mu}),   
\end{equation}
so that the gauge field $A$ can be easily integrated out. Since the field 
$h$ is purely transverse, integer variables 
$\{m_{n\mu}\}$ can be taken to be 
transverse as well, and 
\begin{equation}
Z=lim_{t\rightarrow 0} \int'_h \sum'_m \exp(i 2\pi \sum m_{n\mu} h_{n\mu} - 
\frac{1}{2K_0}\sum(\nabla\times h )_{n\mu}^2
- \frac{e^2}{2} \sum h_{n\mu}^2 -\frac{t}{2} \sum m_{n\mu}^2),  
\end{equation}
where we added a chemical potential term. 
Modulo Villain approximation, this is the same as 
\begin{equation}
Z=lim_{t \rightarrow 0} \int'_{h} \int_{\theta} \exp( \frac{1}{t}
\sum cos(\theta_{n\mu}-\theta_{n}-2\pi h_{n\mu}) -\frac{1}{2K_0}\sum(\nabla
\times h)_{n\mu}^2 - \frac{e^2}{2} \sum h_{n\mu}^2). 
\end{equation}
When $e=0$, the dual partition function (5) is identical to the 
$K_0\rightarrow\infty$ limit of theory in (1). 
In this limit therefore,  the transition is at 
$e^2 =e_c ^2=4\pi^2 K_c$, as originally found by Peskin \cite{peskin}. 
Apart from the limit $t\rightarrow 0$, the last expression is completely 
analogous to the partition function for the lattice superconductor, 
eq. 1, except that the dual gauge-field $h$ is {\it massive}. 
Finite $t$ adds 
a small chemical potential for the loop variables $\{m_{n\mu}\}$ 
in the eq. 4. This suggests 
that the limit $t\rightarrow 0$ is regular, in the sense 
that the nature of the phase transition 
is not changed by leaving $t$ finite. 
 This is the same assumption as the one made by the authors of 
ref. 11 in their argument for the inverted XY transition in 
lattice superconductor. 
Unfortunately, we have been unable to check directly the validity of this 
assumption.
Rather, we  will adopt it for a moment, and return 
to this question after we derive its consequences. 
For $t\neq 0$, the dual model 
of the lattice superconductor  
should be in the same 
universality class as the continuum theory for a complex scalar 
field representing vortex loops (the disorder field) 
minimally coupled to the massive dual gauge-field 
$\vec{h}$ (with $\nabla\cdot\vec{h}=0$) \cite{kovner}, \cite{kiometzis}:
\begin{equation}
L=\int d^{3}\vec{r} 
[|(\nabla - i \frac{2\pi}{e} \vec{h})\Psi(\vec{r})|^{2} + 
\mu^{2} |\Psi(\vec{r})|^{2} + \frac{b_{0}}{2}|\Psi(\vec{r})|^{4}+ 
\frac{1}{2}\vec{h}^{2}+\frac{1}{2\mu_{h}^{2}}(\nabla \times\vec{h})^{2}]. 
\end{equation}
$\mu^{2}\sim(T_{c0}-T)$, $T_{c0}$ is 
the mean-field transition temperature, the quartic term describes the 
short-range repulsion between the 
vortex loops, and $\mu_h^2=e^2 K_0$.  

Having a continuum version of the dual 
theory, the critical behavior can be studied by standard renormalization 
group methods (RG). 
We adopt the multiplicative 
field-theoretic renormalization and calculate the RG factors perturbatively 
and directly in 3D. The perturbation theory 
is used only to make the reasoning more 
explicit, and is not essential for the main conclusions.
The renormalized dual theory is: 
\begin{equation}
L_{r}=\int d^{3}\vec{r}[ Z_{\Psi} |(\nabla-i \frac{2\pi}{e}\vec{h})\Psi|^{2}
+ m'^{2}|\Psi|^{2}+\frac{b'}{2}|\Psi|^{4}+\frac{1}{2}\vec{h}^{2} + 
\frac{Z_{h}}{2\mu_{h}^{2}}(\nabla\times\vec{h})^{2}].
\end{equation}
Notice that the form of minimal coupling between $\Psi$ and $\vec{h}$ 
is preserved and the same Ward identities hold as if $\vec{h}$ was massles. 
This further implies that the polarization of the field $\vec{h}$ is 
transverse and $\vec{h}^{2}$ term does not get renormalized. To 
the leading 
order in two dimensionless coupling constants 
$\hat{b}_{0}=b_{0}/m$ and $\hat{g_{0}}=(2\pi/e)^{2}(\mu_{h}^{2}/m)$ 
we  obtain:
\begin{equation}
Z_{h}=1+\frac{\hat{g_{0}}}{24\pi}, 
\end{equation}
\begin{equation}
Z_{\Psi}=1-\frac{2\hat{g_{0}}}{3\pi(1+(\mu_{h}/m))}, 
\end{equation}
\begin{equation}
b'=b_{0}-\frac{5}{8\pi}\frac{b_{0}^{2}}{m}- \frac{1}{2\pi}\frac{(2\pi\mu_{h}
/e)^{4}}
{\mu_{h}}. 
\end{equation}
With the renormalized coupling constants  defined as $\hat{g}=(2\pi/e)^{2}
(m_{h}^{2}/m)$, $m_{h}^{2}=\mu_{h}^{2}/Z_{h}$, 
$m=m'/Z_{\Psi}^{1/2}$ and $\hat{b}=\hat{b'}/
Z_{\Psi}^{2}$, the one-loop $\beta$-functions are:
\begin{equation}
\beta_{g}\equiv\frac{d\hat{g}}{d \log m}=
-\hat{g}+\frac{1}{24\pi}\hat{g}^{2}, 
\end{equation}
\begin{equation}
\beta_{b}\equiv\frac{d\hat{b}}{d\log m}
=-\hat{b}+\frac{5}{8\pi} \hat{b}^{2}-\frac{2}{3\pi}\frac{\hat{b}
\hat{g}(2+m_{h}/m)}{(1+(m_{h}/m))^2}+ 
\frac{1}{2\pi} \hat{g}^{2} \frac{m}{m_{h}}.
\end{equation}
First, there are two unstable fixed points with 
vanishing dual charge $\hat{g}_{c}
=0$: 
$\hat{b}_{c}=0$ and $\hat{b}_{c}=8\pi/5$. Note that the last two terms 
in $\beta_{b}$ depend not only on $\hat{b}$ and $\hat{g}$, but on the 
third dimensionless combination of the coupling constants 
$m/m_h$ as well. 
The phase transition in the theory 
is achieved by tuning the temperature, that is by letting $m\rightarrow 0$, 
with the scaling of $m_h$ determined by $\beta_g$. 
Close to the attractive fixed point with a finite dual charge $\hat{g}=
\hat{g}_{c}\neq 0$: 
\begin{equation}
(\hat{g}-\hat{g}_{c})= const\cdot m^{\beta'_{g}(\hat{g}_{c})}, 
\end{equation}
so that, 
\begin{equation}
\frac{(2\pi)^2}{e^2} m_{h}^2=const\cdot m^{1+\beta'_{g}(\hat{g}_c)} 
+ \hat{g}_c m. 
\end{equation}
Since $\beta'_{g}(\hat{g}_{c})=1>0$, when $m \rightarrow 0$ 
the renormalized stiffness constant $K =m_{h}^2/e^2$ 
close to the transition approaches zero as
\begin{equation}
K =\frac{\hat{g}_c}{(2\pi)^2} m.  
\end{equation}
Recalling that $m\equiv \xi^{-1}$, where $\xi$ is the correlation length 
of the disorder field, we see that 
the result is in accordance with the Josephson 
relation in 3D \cite{joseph}. To the lowest order in dual charge, the 
universal amplitude is 
\begin{equation}
lim_{m\rightarrow 0}\frac{K}{m}=\frac{6}{\pi}. 
\end{equation} 
Now it becomes obvious that the last two terms 
in $\beta_{b}$ vanish as the critical temperature is approached 
because $m_h/m\rightarrow \infty$, in spite of the scaling of 
dual charge towards a finite fixed point value. Consequently, the attractive 
fixed point with $\hat{g}_c\neq 0$ is located 
at the same $\hat{b}_c=8\pi/5$ as the pure XY fixed point. 
As may have been expected, the fluctuations of the 
massive dual gauge-field do not influence the critical behavior of 
the theory, which is in the universality class of 
3D XY model \cite{kovner}, \cite{kiometzis}. 
This is in sharp contrast with the effects of the 
massles gauge-potential in the Ginzburg-Landau theory \cite{halperin}, 
\cite{herbut}, \cite{bergerhof}.
We proved this statement here only to the lowest order, but 
clearly it must persist 
to all orders in perturbation theory. One could integrate out the dual gauge 
field $\vec{h}$ from the beginning and  the only effect of that (apart from 
introducing some irrelevant terms) would be the 
modification of the bare quartic term for the disorder 	field. 

From the partition function for the 
lattice superconductor, eq. 1, it follows that the magnetic 
field penetration depth $\lambda$ 
scales with the stiffness constant as :
\begin{equation}
\lambda^{-(2-\eta_{A})}\propto K, 
\end{equation}
where $\eta_{A}$ is the anomalous dimension acquired by 
the gauge-field at the 
stable fixed point in the original theory. At the neutral, unstable fixed 
point with $e=0$, $\eta_{A}=0$. At 
the charged, attractive fixed point in the theory however, 
$\eta_{A}=1$ in 3D \cite{herbut}. This may also be understood by 
recalling that in the continuum limit the dimension of the 
charge $[e^2]= L^{-1}$, so that 
close to the stable fixed point scaling of the 
$(\nabla\times A)^2$ term with length is changed by one power. 
From the relations  (15) and (17) it follows that 
\begin{equation}
\lambda\propto \xi, 
\end{equation}
and the exponents for the correlation length 
of the disorder field and for the magnetic field penetration depth 
are the same. 
This result is independent of the perturbation theory.  
It was only necessary that 
$\beta'_{g}(\hat{g}_c)>0$, which is always true at a non-trivial 
zero of $\beta_g$. The continuum dual theory (6) thus implies that 
the exponent for the magnetic field penetration depth 
has the XY value $\nu_{xy}\simeq 0.67$. 

The same form of the continuum 
dual theory (6)  has previously been 
considered in ref. 13. 
However, by making some phenomenological assumptions for the coupling 
constants 
in the theory, these authors obtained an incorrect value for the 
exponent for the penetration depth. The essential difference with the 
present work is 
that the bare mass of the dual gauge field $\vec{h}$ was assumed to 
independently vanish close to the critical point as $\mu_{h}^{2}
\propto (T_c -T) \propto m^{x}$, 
 $x=1/\nu_{xy}$. 
The lowest order $\beta$-function for the dual charge 
is then modified to read:
\begin{equation}
\beta_{g}=(x-1)(\hat{g}-\frac{1}{24\pi}\hat{g}^{2}), 
\end{equation}
and the neutral ($\hat{g}_{c}=0$) fixed point of the dual theory 
becomes infrared stable 
for any $x>1$. 
Since $\beta'_{g}(0)=x-1$, eq. 13 then implies that 
\begin{equation}
m_{h}^{2}\sim m^{x}.  
\end{equation}
The power $x$ does not get changed by the fluctuations of the 
disorder field if its assumed value is larger than unity. 
The authors of ref. 13 interpreted this observation as that 
the exponent for the penetration depth has the mean-field value $1/2$. 
It is obvious however, that any other assumption for the power $x$ would, 
as long as $x>1$, 
lead to a correspondingly  different exponent 
for the penetration depth. Furthermore,  
the interpretation of the field theory (6) as a 
continuum limit of the dual lattice superconductor makes it 
clear that nothing should be assumed for the mass $\mu_h$. 
Close to the transition, the renormalized mass $m_h$ 
is then {\it driven} to zero by 
fluctuations of the disorder field, as a power law 
that follows from the theory. 

The behavior of the penetration depth derived here on the basis of 
the continuum dual theory is in disagreement with of results 
obtained from the Ginzburg-Landau theory for the superconducting 
order-parameter. Perturbative 
\cite{herbut} and non-perturbative \cite{bergerhof} renormalization 
group calculations 
in fixed dimension indicate that the penetration depth in Ginzburg-Landau 
model 
diverges with the exponent $0.50<\nu<0.62$. The main point is that 
the exponent obtained from Ginzburg-Landau theory is 
different (and most likely smaller) than $\nu_{xy}$. In fact, this 
is what one would naively expect, since the neutral, XY fixed point 
in the Ginzburg-Landau 
theory is unstable with respect to the charge, and small perturbations 
in that direction are attracted by the stable, charged fixed point 
with different exponents.
One conceivable reason for this disagreement is the 
approximate nature of the calculations performed in refs. 9 and 10. 
One should note however,  that the non-perturbative method employed 
in ref. 10 leads to very accurate values for the exponents in several 
other problems, where there is more information available. 
A more interesting possibility is that the condition 
$t\rightarrow 0$ which we were forced to relax in order to arrive at the 
continuum theory, might represent a singular limit in 
the theory \cite{remark}. In the eq. 4, for $t=0$, the summation over 
integers $\{m_{n\mu}\}$ would force the dual 
gauge field $h$ to take strictly integer 
values. Finite $t$ relaxes this constraint, and $h$ becomes distributed 
around integers with a finite width of distribution proportional to $t$. 
When viewed this way, it becomes less clear that finite and zero $t$ 
necessarily lead to the same physics. More detailed numerical studies should be 
able to shed some light on the problem. 

 The divergence of the penetration depth close to the 
superconducting transition is experimentally measurable, at least 
in principle. In practice, unfortunately, this is made prohibitively difficult 
by smallness of the critical region for the gauge-field fluctuations, 
which is rather narrow (even in the high-temperature superconductors) due 
to small value of the effective charge (Ginzburg-Landau parameter 
$\kappa >> 1$). The presently 
accessible critical region corresponds to the vicinity of the unstable 
neutral fixed point in the Ginzburg-Landau theory, at which the 
penetration depth diverges with the exponent $\nu_{xy}/2$ (because 
$\eta_{A}=0$ at this fixed point)
\cite{bonn}. We should note here however, that 
there are other physical systems, where the 
critical region for the gauge-field fluctuations is much wider \cite{degen}, 
\cite{zlatko} and which offer more hope for the experimental resolution 
of the question of the exponents at the charged critical point.

In summary, the message of this paper is twofold. 
First, we studied the phase transition in 3D lattice superconductor 
within a continuum version of the dual theory which yields the  
transition to be in the XY universality class. 
The superfluid density  vanishes as inverse 
of the correlation length of the disorder field. In contrast 
to the previous claims, dual charge is a relevant coupling whose 
fixed point value determines the universal amplitude for the 
scaling of the stiffness constant. This amplitude is calculated to the 
lowest order in perturbation theory. Finally, we have shown that 
within the continuum dual theory the 
magnetic field penetration depth diverges 
with the exponent $\nu_{xy}\simeq 2/3$. Second, it is pointed out that 
the critical behavior obtained on the basis of the continuum dual 
theory is in disagreement with the recent results from 
Ginzburg-Landau theory for the superconducting order-parameter. Possible 
reasons for this disagreement are discussed and experimental consequences 
are mentioned. 

Many useful discussions with Professors Z. Te\v sanovi\' c and I. Affleck 
are greatfully acknowledged. 
This research has been supported by Natural Sciences and 
Engineering Research Council of Canada and Izaak Walton Killam 
foundation.

%\pagebreak


\begin{thebibliography}{99}
\bibitem{remark} This would agree with the ref. 14, where 
similarly, the "frozen" 
($K_0\rightarrow\infty$) limit of lattice superconductor was considered.
\end{thebibliography}
\end{document}